\documentclass[doublecol,linenumbers]{epl2} 

\usepackage{graphicx}
\usepackage{epsf}
\usepackage{dcolumn}
\usepackage{bm}
\usepackage[utf8]{inputenc}
\usepackage{amsmath}
\usepackage{xcolor}

\newcommand\R[1]{\mathrm{#1}}

\title{Correlations in the chaotic spectrum of pressure modes in rapidly rotating stars}

\author{Benjamin Evano\inst{1,2} \and Bertrand Georgeot\inst{2} \and Fran\c{c}ois Ligni\`{e}res\inst{1}}
\shortauthor{B. Evano \etal}

\institute{                    
  \inst{1} IRAP, Université de Toulouse, CNRS, CNES, UPS, (Toulouse), France\\
  \inst{2} Laboratoire de Physique Théorique, IRSAMC, Université de Toulouse, CNRS, UPS, France
}


\abstract{The oscillation spectrum of pressure waves in stars can be determined by monitoring their luminosity. For rapidly rotating stars, the corresponding ray dynamics is mixed, with chaotic and regular zones in phase space. Our numerical simulations show that the chaotic spectra of these systems exhibit strong peaks in the autocorrelation which are at odd with Random Matrix Theory predictions. We explain these peaks through a semiclassical theory based on the peculiar distribution of the actions of classical periodic orbits. Indeed this distribution is strongly bunched around the average action between two consecutive rebounds and its multiples. In stars this phenomenon is a direct consequence of the strong decrease of the sound speed towards the star surface, but it would arise in any other physical system with a similar bunching of orbit actions. The peaks discussed could be observed by space missions and give insight on the star interiors.}

\begin{document}

\maketitle

\section{Introduction} Most of the information that we can obtain from stars stems from the light that they emit. Variations of this light can be monitored for many stars, enabling to detect periodic patterns produced by oscillation modes of stars. In the case of the Sun, it has been possible to theoretically construct these modes and compare them with observations, giving crucial pieces of information on the internal structure \cite{christensen-dalsgaard_helioseismology_2002}. Ultra-precise photometric data from the recent space missions COROT \cite{baglin200636th} and Kepler \cite{borucki_kepler_2010} include many rapidly rotating stars, for which a theory of oscillation modes is needed to infer physical properties of their internal structure \cite{aerts_asteroseismology_2010}.
 
Stellar oscillations are divided into two categories that are separated in frequencies, with gravity modes below the Brunt-Väisälä frequency and pressure (acoustic) modes in the high part of the frequency spectrum. Recently, ray dynamics and semiclassical techniques have been used to describe oscillation modes in rapidly rotating stars \cite{lignieres_wave_2008, prat_asymptotic_2016}. Indeed, these oscillations have a short-wavelength limit in the same way as quantum or electromagnetic waves do, and the ray dynamics is also governed by Hamiltonian equations of motion \cite{gough_linear_1993}. In the case of acoustic waves, which are the focus of this letter, numerical simulations of this dynamics for a polytropic stellar model showed that when the rotation rate increases, stable and chaotic regions coexist in phase space. As a consequence, the stationary acoustic modes can be divided in modes localized in the regular zones or in the chaotic zones, with markedly different properties \cite{lignieres_wave_2008}. For regular island modes an asymptotic formula was built in \cite{pasek_regular_2011}, showing that they are characterized by regular spacings. Chaotic modes, which have been studied in the context of quantum mechanics by the field of quantum chaos \cite{gutzwiller_chaos_1990}, are expected to be distributed in accordance with the predictions of Random Matrix Theory (the relevant matrix ensemble for our system is the Gaussian Orthogonal Ensemble (GOE)) \cite{bohigas_characterization_1984, bohigas_houches_1989}, and should not display regularities.

In this Letter we show that chaotic modes in models of rapidly rotating stars display pseudo-regularities which can be clearly seen from peaks in the autocorrelation of the spectra. They could correspond to peaks extracted from the observed frequency spectra of $\delta$ Scuti stars \cite{garcia_hernandez_observational_2015}, a class of rapidly rotating pulsating stars. Understanding their origin would then be key to derive physical constraints on star interiors. These peaks are also of interest from a theoretical standpoint, since they are unseen in the autocorrelation of GOE spectra \cite{bohigas_houches_1989}. Here, we explain the pseudo-regularities in the chaotic mode spectrum from a theory based on the general properties of acoustic rays in stars, coupled with the semiclassical theory of correlations in chaotic spectra \cite{berry_semiclassical_1985}.

\section{Pseudo-regularities in the chaotic spectrum of stars} To model the star, we study a self-gravitating monoatomic perfect gas of adiabatic exponent $\Gamma = 5/3$ in solid body rotation. The equilibrium model is a simplified model, where the pressure and density satisfy the relation $P \propto \rho^{1+1/N}$ with an effective polytropic index N=3. The acoustic ray dynamics is derived making standard approximations valid for high frequencies : we neglect the Coriolis force, the perturbations of the gravitational potential and the effect of viscosity and thermal diffusivity. In the linear limit, one obtains a Helmholtz-type equation. The pressure perturbations $\Psi(\boldsymbol{x},t)$ satisfy :

\begin{equation}
    \left( \frac{\partial^2}{\partial t^2} + \omega_c^2 \right) \Psi(\boldsymbol{x},t) - c_s^2 \, \nabla ^2 \Psi(\boldsymbol{x},t) = 0
        \label{eqac}
\end{equation}

\noindent where $c_s$ is the sound speed, that decreases from the center to the surface as the square root of the temperature and $\omega_c$ is a cutoff frequency which increases sharply close to the boundary, confining the wave inside the star \cite{lignieres_asymptotic_2009}.

We seek solutions of the form $\Psi = e^{i \Lambda \phi(\boldsymbol{x},t)}$ where $\Lambda^{-1}$ is a small dimensionless parameter and get an equation for the phase $\phi(\boldsymbol{x},t)$ \cite{gough_linear_1993} :

\begin{equation}
    -\Lambda \dot{\phi}(\boldsymbol{x},t) = (\omega_c^2 + c_s^2 k^2)^{1/2} = H
    \label{H-J}
\end{equation}

\noindent where $\boldsymbol{k} = \boldsymbol{\nabla}{\phi}$ is the wavevector and $H$ is the ray dynamics hamiltonian. As is usual in stellar physics \cite{aerts_asteroseismology_2010}, rotation frequencies will be given in units of the Keplerian rotation rate $\Omega_k = (GM/R_{eq}^3)^{1/2}$ where the centrifugal acceleration equates gravity at the equator, $G$ being Newton's constant, $M$ the mass and $R_{eq}$ the equatorial radius of the star. We will express acoustic frequencies in terms of $\omega_p = (GM/R_p^3)^{1/2}$, where $R_p$ is the polar radius. Because $R_p$ varies slowly with rotation, the choice of $\omega_p$ is well suited to compare acoustic frequencies at different rotation rates. \\

\begin{figure}
\onefigure[width = 1\columnwidth]{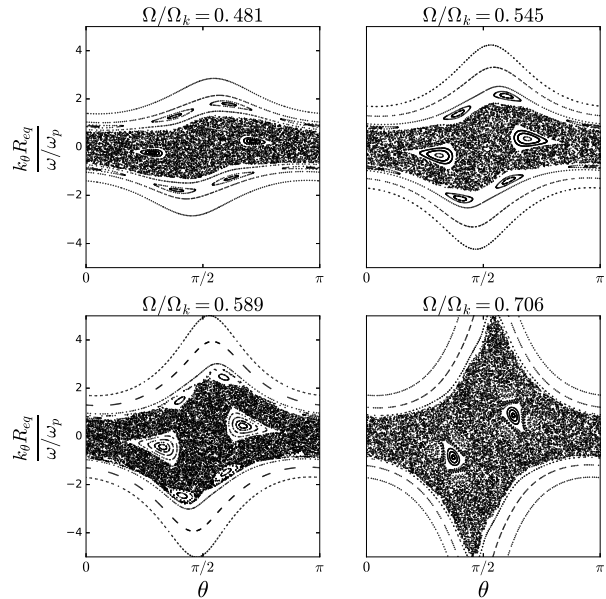}
\caption{PSS of acoustic rays at four different rotations. Each point corresponds to the crossing of an outgoing ray with a curve at constant distance from the star surface, $\theta$ being the colatitude and $k_{\theta}$ the wavevector component parallel to the curve at the crossing point.}
\label{PSS_evolution}
\end{figure}

The equations of motion are solved using a 5th order Runge-Kutta method. As the system is cylindrically symmetric, the ray dynamics can be reduced to a two-dimensional problem \cite{lignieres_asymptotic_2009}. In the particular case of axisymmetric modes, acoustic rays propagate in a meridional plane of the star. The study of the dynamics through the Poincar\'e Surface of Section (PSS) reveals three main types of structures for a wide range of rotation rates (see fig.~\ref{PSS_evolution}) : stable islands are built around periodic orbits, whispering gallery rays remain close to the surface and chaotic ergodic trajectories fill all available space. Stationary modes are associated to these different phase space regions. As presented in \cite{lignieres_asymptotic_2009}, modes associated with the stable islands have simple spectra of the form $\omega_{n \ell} = n \delta n + \ell \delta \ell + \alpha$ and in the case of the most important 2-period island modes this formula was explicited by an asymptotic theory \cite{pasek_regular_2011}.\\

\begin{figure}
\onefigure[width = 1\columnwidth]{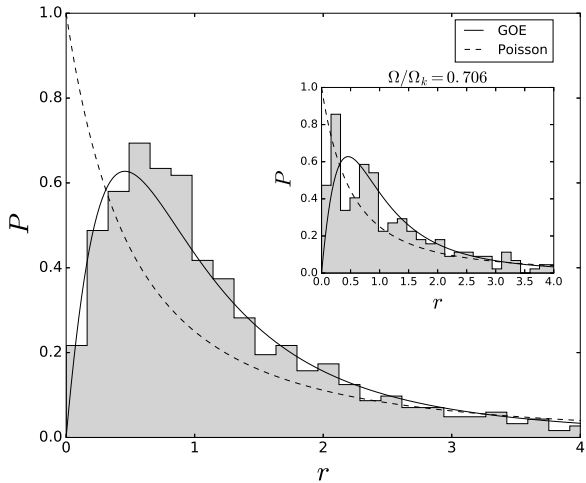}
\caption{The ratio distribution $P(r)$, with 1174 levels from seven independent spectra : $\Omega / \Omega_k = 0.481$ (206 odd levels), $\Omega / \Omega_k = 0.545$ (223 odd levels, 105 even levels), $\Omega / \Omega_k = 0.589$ (217 odd levels, 96 even levels), $\Omega / \Omega_k = 0.658$ (207 odd levels, 120 even levels). The solid line is the GOE distribution and the dashed line is the Poisson distribution. The special case $\Omega / \Omega_k = 0.706$, with 283 odd levels, is shown in the inset.}
\label{p(r)}
\end{figure}

By contrast, chaotic modes associated with the ergodic phase space region are not predicted in general to follow any simple asymptotic formula. According to the Bohigas-Giannoni-Schmit conjecture \cite{bohigas_characterization_1984} chaotic mode spectra should have correlations given by GOE, that have the property of repelling each other at short distance. Conversely, generic regular modes should follow a Poisson distribution with no level repulsion \cite{berry_level_1977}. Using a two-dimensional code that computes the stationary oscillations of a rotating star \cite{reese_acoustic_2006} we produced the frequency spectra of the polytropic model at five different rotations (namely the four rotations shown in fig.1 and additionally $\Omega/\Omega_k$ = 0.658, whose PSS is very similar to panel c but with a disappearing 6-periodic island chain). All modes are axisymmetric and either symmetric (even) or antisymmetric (odd) with respect to the equator. We then selected the chaotic modes by removing from the full spectra the island modes and whispering gallery modes \footnote{The island modes where removed by hand using the known formula for their frequencies as a guide. The whispering gallery modes were removed automatically using the fact that their spherical harmonic decomposition is dominated by large degree components.}. In fig.~\ref{p(r)} we display the distribution of the ratio of consecutive level spacings $r_n = (\omega_{n+1}-\omega_n)/(\omega_n-\omega_{n-1})$ of chaotic modes and in table \ref{ratio_mean} we indicate the average value of $\tilde{r}_n = \mathrm{min}(r_n, 1/r_n)$ \cite{oganesyan_localization_2007, atas_distribution_2013}. The main panel of fig.\ref{p(r)} shows the aggregated ratio distributions of four rotations. This procedure is standard to smooth out the statistical fluctuations due to small datasets \cite{bohigas_nuclei}. The resulting distribution is in good agreement with the GOE prediction for chaotic systems. This is not the case for the $\Omega / \Omega_k = 0.706$ spectrum, shown in the inset, as the ratio distribution does not show level repulsion at small $r$ values and the averaged $\tilde{r}$ falls in between the GOE and Poisson values. Such anomalous statistics is often associated to the fact that the spectrum contains independent subspectra. We will explain this peculiarity below by the presence of partial barriers in the chaotic zone.

\begin{table}
\caption{Average value of $\tilde{r}$.}
\label{ratio_mean}
\begin{center}
\begin{tabular}{rl}
 \hline
 Poisson &$\approx 0.38629$ \\
 GOE  &$\approx 0.53590$ \\
 $\Omega / \Omega_k = 0.706$ &$\approx 0.4743$ \\
 Other rotations &$\approx 0.5618$ \\
 \hline
\end{tabular}
\end{center}
\end{table}

\begin{figure}
\onefigure[width = 1\columnwidth]{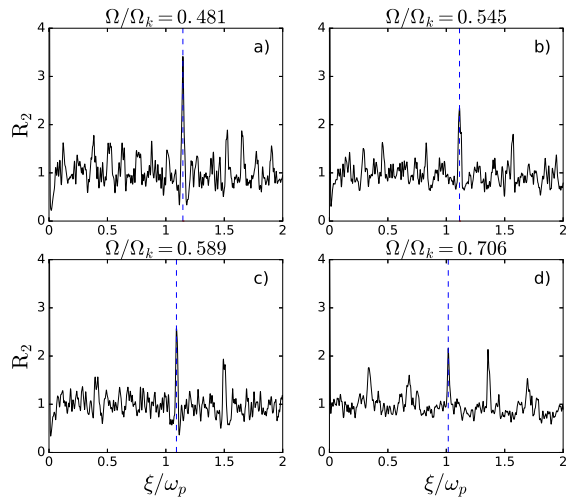}
\caption{Frequency autocorrelations computed at four rotations for chaotic modes, using only one symmetry class (odd modes) since odd and even spectra are independent. Datasets : a) 206 levels from 28.35 $\omega_p$ to 46.89 $\omega_p$, b) 223 levels from 28.15 $\omega_p$ to 44.09 $\omega_p$, c) 217 levels from 26.02 $\omega_p$ to 40.29 $\omega_p$ and d) 283 levels from 23.57 $\omega_p$ to 36.22 $\omega_p$. Dashed lines mark the position of the most robust peak, at much longer range than the mean level spacing ($\approx 0.09 \, \omega_p$).}
\label{Auto}
\end{figure}

The ratio of consecutive level spacings is a short-range quantity; to investigate the correlations at longer range we computed the two-point autocorrelation function $\R{R_2}(\xi) = \left<\R{d}(\omega-1/2 \xi) \, \R{d}(\omega+1/2 \xi) \right>$, where $\R{d}(\omega)$ is the spectral density and $\left<.\right>$ is a frequency average defined by $\left< f \right> = \int f \R{d}\omega$. As shown in fig.~\ref{Auto}, a deviation from GOE statistics \cite{bohigas_houches_1989} appears in the form of a peak centered at value $ \Delta \simeq 1.15 \, \omega_p$ (panel a). This peak is robust in the sense that it appears at every rotation rate, though its position shifts slightly (see fig.~\ref{Auto} panels b, c, d). Other peaks emerge from the noise, especially as rotation increases (fig.~\ref{Auto}, panels c and d) around $1/3 \Delta$ and $4/3 \Delta$. In the following, we will explain the origin of the most robust peaks and their evolution with respect to rotation, using semiclassical methods. We will also link the presence of the additional peaks to the existence of partial barriers.

\section{Semiclassical analysis} Semiclassics (here large $\omega$ limit) is built on the Gutzwiller trace formula, which relates the spectral density to a sum over periodic orbits of the system \cite{gutzwiller_chaos_1990}, and was introduced in the framework of quantum mechanics. We have adapted the derivation to the case of acoustic modes; it leads to the formula $\R{d}(\omega)-\R{\bar{d}}(\omega) = \mathrm{Re}\sum_i A_i(\omega) \, e^{\R{i} S_i (\omega)}$ where $i$ labels periodic orbits, $S$ is the action, $A$ is an amplitude, supposed to be slowly varying with $\omega $, which depends on the length and stability of the orbit and $\R{\bar{d}}(\omega)$ is the smoothed density. In our system the action can be expressed as $S = \omega \displaystyle \int \R{d}s \, \sqrt{1-\omega_c^2 / \omega^2} / c_s$, where $s$ is the curvilinear coordinate along a given ray path. Therefore $T = \partial S / \partial \omega$ corresponds to the travel time (or acoustic time) of a ray.

\noindent In order to relate the autocorrelation of the chaotic spectra to the acoustic ray dynamics, we follow the method of Berry \cite{berry_semiclassical_1985}, which uses the so-called diagonal approximation. The idea is to consider the form factor :

\begin{equation}
    \R{K}(T) = \frac{1}{\sqrt{2 \pi}}\int^{\infty}_{-\infty} \R{d} \xi \, \R{exp}(\R{i} \xi T) \, \R{C}(\xi)
    \label{form_factor}
\end{equation}

\noindent which is the Fourier transform of the autocorrelation :

\begin{equation}
\begin{aligned}
\R{C}(\xi) &=  & \left<[\R{d}(\omega-1/2 \xi)-\R{\bar{d}}(\omega-1/2 \xi)]\right. \times \\
           &   & \left. [\R{d}(\omega+1/2 \xi)-\R{\bar{d}}(\omega+1/2 \xi)] \right> \\
           &=  & \Big \langle \Big(\R{Re} \sum_i A_i e^{\R{i} S_i(\omega - 1/2 \xi)}\Big) \times \\
           &   & \Big(\R{Re} \sum_j A_j e^{\R{i} S_j(\omega + 1/2 \xi)}\Big) \Big \rangle
\end{aligned}
\end{equation}

\noindent For short times, the frequency average eliminates the off-diagonal terms $i \neq j$ by phase incoherence, allowing to approximate the form factor as $\R{K}(T) \approx \sum_j A_j^2 \delta(T-T_j)$, \textit{i.e.} the density of periodic orbits weighted with intensities $A_j^2$ \cite{berry_semiclassical_1985, berry_houches_1989}. This approximation fails for long times, typically higher than the Ehrenfest time \cite{bogomolny_arithmetic}, because there are pairs of orbits with very close actions $S_i \approx S_j$ and higher order terms need to be computed \cite{bogomolny_gutzwillers_1996, sieber_correlations_2001, heusler_periodic-orbit_2007, muller_periodic-orbit_2009}.

\begin{figure}
\onefigure[width = 1\columnwidth]{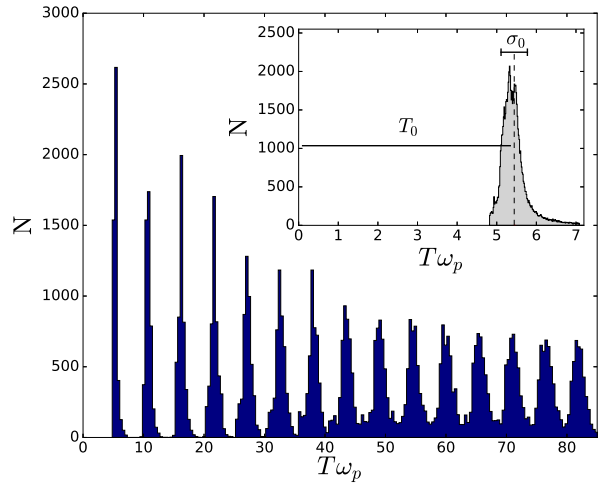}
\caption{Number of n-chord trajectories, with n = 1, ..., 15, vs their travel time $T$ at rotation $\Omega / \Omega_k = 0.589$. The n-chord samples contain 4750 chords each. The inset shows the 1-chord distribution in more details, with $70000$ chords. The dashed line marks its mean value, very close to $T_0$.}
\label{histo}
\end{figure}

The next step of Berry's method is to use the general result of Hannay and Ozorio de Almeida \cite{hannay_periodic_1984} which implies that $\sum_j A_j^2 \delta(T-T_j) \propto T$ for sufficiently large $T$. To make this result explicit we express the amplitude and the density as functions of the metric and topological entropies $h_M$, $h_T$ : $A (T) \propto T e^{- (1/2) h_M T}$ and $\sum_j \delta(T-T_j) \rightarrow \rho(T) \approx (1/T) \, e^{h_T T}$ and suppose these two entropies to be equal \cite{adler_topological_2008, berry_houches_1989}. These formulas give: $\R{K}(T) \rightarrow A^2(T) \; \rho(T) \propto 2 T$, which is the GOE prediction at leading order (the factor of two comes from the fact that orbits are twice degenerate due to time-reversal symmetry \cite{berry_semiclassical_1985}). In our system, however, the peculiar distribution of periodic orbits $\rho(T)$ modifies the form factor behavior.

\par
Like classical trajectories in billiards, acoustic rays bounce on the reflective caustic close to the surface. Thus any periodic orbit can be divided into chords, each chord connecting two successive surface bouncing points. Periodic orbits thus belong to a class of trajectories that we call n-chords, whose endpoints have to lie on the surface. While it is extremely difficult to find all periodic orbits of our system, we can nevertheless infer some of their properties from the study of large samples of n-chord trajectories.

\begin{figure}
\onefigure[width = 1\columnwidth]{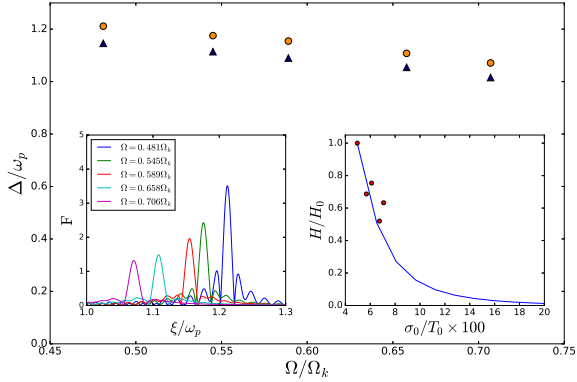}
\caption{Theoretical value of $\Delta/\omega_p$ (circles) compared to the numerical peak's position (triangles) for five rotation rates. Left inset : theoretical autocorrelations around the main peak, from left to right the rotation rate decreases. Right inset : the solid line shows the variation of the normalized peak height $H/H_0$, where $H_0$ is the peak height at $\Omega/\Omega_k = 0.481$, with increasing $\sigma_0$ at fixed  $T_0$. The points show the corresponding values obtained from the simulated spectra.}
\label{results}
\end{figure}

Fig.~\ref{histo} shows the distribution of travel times for n-chords in the chaotic region of phase space at rotation $\Omega / \Omega_k = 0.589$. In stars, the travel time is not proportional to the geometric length, as \textit{e.g.} in billiards. Indeed, since the sound velocity is much smaller near the surface of the star, all trajectories spend much more time near the surface than in the core. This results in a travel time distribution that concentrates around specific values evenly spaced out (see fig.~\ref{histo}). As the number of chords increases the individual packets grow wider (similarly to the law of large numbers),  until the width becomes much larger than the interpeak distance, and the distribution turns into a smooth curve. This is quantified by the ratio $\sigma_0/T_0$, $\sigma_0$ being the standard deviation of the first packet and $T_0$ the average distance between the mean values of consecutive packets. In the case of a billiard whose boundary is shaped like our star surface, $\sigma_0/T_0 \approx 0.32$ meaning that the packet structure disappears after a few rebounds. In the stellar case $\sigma_0/T_0$ grows with rotation but remains below $0.08$. The packets are thus discernable for much longer times. \\

\begin{figure}
\onefigure[width = 1\columnwidth]{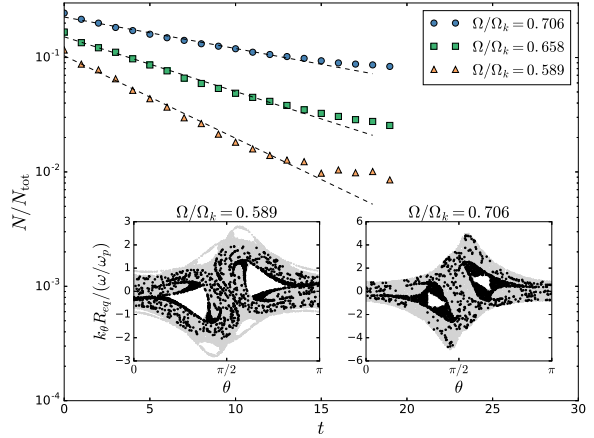}
\caption{Monte Carlo simulations showing $N/N_{tot}$ vs $t$ for different rotations, where $t$ is a discretized time (number of crossing of rays with the PSS), $N_{tot}$ the number of initial chaotic trajectories and $N$ the number of trajectories that remain in the partial barrier zone. The intercept is representative of the relative area and the slope gives the escape rate $\lambda$, \textit{i.e.} the outward flux : $ \lambda(\Omega / \Omega_k=0.589) \simeq 0.17$, $\lambda(\Omega / \Omega_k =0.658) \simeq 0.11$ and $ \lambda(\Omega / \Omega_k = 0.706)\simeq 0.06$. Insets are snapshots of the time evolution showing the zones delimited by partial barriers.}
\label{barriers}
\end{figure}

We model the distribution of fig.~\ref{histo} as $\R{P}(T) = \sum_n \R{P_n}(T)$, with $\R{P_n}(T)$ the probability distribution of n-chords travel times :

\begin{equation}
    \R{P_n}(T) = \frac{1}{\sqrt{2 \pi n} \, \sigma_0} \, \R{exp} \left( - \frac{(T-n \, T_0)^2}{2 (\sqrt{n} \, \sigma_0)^2} \right)
    \label{podist}
\end{equation}

\noindent where the values of $T_0$ and  $\sigma_0$ are taken from the numerical simulations of acoustic rays. We checked that the packets have indeed a mean value close to $nT_0$ and that their standard deviation grows as $\sqrt{n}\sigma_0$. Physically, $T_0$ is the characteristic travel time of a single chord. Thus $T_0$ and $\sigma_0$ are directly related to the star volume and the sound speed profile in the star interior. In particular, the small variance of the travel times is a consequence of the strong decrease of the sound speed near the surface. As mentioned before we cannot find directly the periodic orbits, but we claim that the constraints imposed by $\R{P}(T)$ are strong enough to explain the correlations seen in the spectra. \\

To take into account the exponential growth rate of the number of periodic orbits with $T$, the density of periodic orbits in our system is thus modeled as $\rho(T) \propto (1/T) \, e^{h_T T} \times \R{P}(T)$. Assuming $A(T) \propto T e^{-(1/2)h_M T}$ gives the result

\begin{equation}
\R{K}(T) \rightarrow A^2(T) \; \rho(T) \propto  T \; \R{P}(T)
       \label{theory}
\end{equation}

As the variance of the $\R{P_n}(T)$ grows with $n$, $\R{P}(T)$ becomes flat for sufficiently long times. For such long times we recover the sum rule \cite{hannay_periodic_1984} $\sum_j A_j^2 \delta(T-T_j) \propto T$, which is consistent with the existence of GOE statistics at short spectral distance (fig.~\ref{p(r)}). For shorter times $\R{P}(T)$ creates a specific regime departing from GOE.\par
The Fourier transform

\begin{equation*}
    F(\xi) = 1/(\sqrt{2 \pi}) \int_{-\infty}^\infty \R{d}T \, \exp(-i \xi T) \, T \R{P}(T)
\end{equation*}

\noindent is shown in the left inset of fig.~\ref{results} for different rotations. It shows a peak, as in the spectral data (fig.~\ref{Auto}), that moves towards low frequencies for increasing rotation. The peaks extracted from the numerical spectral data and obtained from semiclassical theory are in good agreement (fig.~\ref{results}, main panel). The discrepancy of about $5 \%$ is similar to what was obtained for island modes in \cite{pasek_regular_2011}, and can be attributed to the relatively low values of numerically computed frequencies, as the theoretical values are derived under the semiclassical (high frequency) approximation. The model also predicts that the peak width should show a slow increase with rotation, as can be seen in the left panel of fig.\ref{results}. In the right panel, we show that the peak height decreases rapidly with $\sigma_0/T_0$; this explains why the peak is not usally visible in chaotic systems with large $\sigma_0 / T_0$.

\section{Other peaks} The semiclassical theory outlined above explains the origin of the main peak in the autocorrelation. Other peaks are sometimes visible in the autocorrelations of fig.~\ref{Auto}, especially for the rotation $\Omega / \Omega_k = 0.706$ where $P(r)$ deviates strongly from GOE (see fig.~\ref{p(r)}). We attribute these peaks to the presence of partial barriers in the chaotic zone around the main stable island. Partial barriers isolate some zones of phase space, from which trajectories escape more slowly than in the rest of the chaotic zone \cite{bohigas_manifestations_1993, shim_whispering_2011}. To find them we compute trajectories with initial condition near the main island and monitor their evolution on the PSS. Without barrier all points would spread ergodically, however fig.~\ref{barriers} shows that a subset of points remains near the island for a long time. Close to $\Omega / \Omega_k = 0.706$, the region enclosed by partial barriers grows in size and traps trajectories more efficiently. For the relatively low frequency waves here considered, they will act as barriers and quantize independently a subset of modes. This will weaken the level repulsion at short spectral distance as seen in fig.~\ref{p(r)}. Besides, as trapped trajectories revolve around a 6-periodic orbit, we expect the modes to quantize like island modes \cite{pasek_regular_2011}, leading to a $\Delta/3$ regularity.

\section{Conclusion} In this Letter, we have shown that specificities in the distribution of the actions of periodic orbits can create a peak in the autocorrelation of chaotic spectra. Such a peak can potentially be detected in the frequency spectrum of chaotic pressure modes in rapidly rotating stars. Recent data have confirmed the existence of peaks in the autocorrelation spectrum of the rapidly rotating $\delta$ Scuti stars \cite{garcia_hernandez_observational_2015}. They have been attributed so far to regular island modes but our results indicate that chaotic modes should also produce such peaks. These two kinds of peaks should be close \footnote{The position of the peak created by island modes is the inverse of the travel time along a specific regular trajectory at the center of the island \cite{pasek_regular_2011}. The peak of chaotic modes corresponds to the inverse of a mean travel time between two rebounds for chaotic trajectories. In view of the 1-chord distribution shown in the inset of fig. \ref{histo} these two quantities should be close, which is confirmed by frequency computations.} but still distinct at most rotation rates. Wheter these two peaks could be discernable in observed data remains to be investigated. In addition to astrophysical observations, the phenomenon described here could be tested with an experimental setup such as \textit{e.g.} electromagnetic waves in a cavity with a strong gradient of the refractive index along the radial direction.

\acknowledgments
We thank CALMIP for providing computational resources and the Universit\'e Paul Sabatier for funding. This work has made use of the TOP code developped by D. Reese and of a user-friendly version of this code developped by B. Putigny.

\end{document}